\documentstyle[preprint,aps,prb,epsf]{revtex}
\begin{document}
\title{An Electron Paramagnetic Resonance Study of Electron-Hole Asymmetry in Charge Ordered Pr$_{1-x}$Ca$_x$MnO$_3$ (x=0.64,0.36)}
\author{ Janhavi P. Joshi, A. K. Sood and S. V. Bhat}
\address{
Department of Physics,
Indian Institute of Science,
Bangalore 560 012, India
}
\author{ K. Vijaya Sarathy and C. N. R. Rao}
\address{
Chemistry and Physics of Materials Unit, Jawaharlal Nehru Centre for Advanced Scientific Research, Jakkur P.O, Bangalore 560 064, India}
\maketitle
\widetext
\begin{abstract}

 We present and compare the results of temperature-dependent electron paramagnetic resonance (EPR) studies on Pr$_{1-x}$Ca$_x$MnO$_3$ for x = 0.64 which is  electron-doped with those on the hole doped x = 0.36 composition. The temperature dependence of the various parameters obtained from the powder and single crystal spectra  show  significant differences between the two manganites. At room temperature the `g' parameter for the electron doped system is less than the free electron `g' value `g$_e$', whereas for the hole-doped system it is more than `g$_e$'. Further, the `g' value and the linewidth obtained from the powder spectra as well as the single crystal spectra show  different functional dependences on temperature in the two systems.   Quite strikingly, the peak observed at T$_{co}$ in the temperature dependence of the asymmetry parameter, $\alpha$, of the single crystal spectra in the hole-doped system is absent in the electron-doped system. We understand this contrasting behaviour of the EPR parameters in the two systems in terms of very different nature of microscopic interactions in them.


\end{abstract}

\section {Introduction}

The phenomenon of charge ordering  observed  in Re$_{1-x}$A$_x$MnO$_3$ where Re is a  trivalent rare earth ion and A is a divalent alkaline earth ion has been  intensively studied for the past few years. This has been one of the most puzzling of the various properties exhibited by the rare earth manganites such as Colossal Magneto Resistance  (CMR), orbital ordering, phase separation  and a large number of magnetic and structural transitions as a function of temperature and composition \cite{rao,tok}. The parent compound ReMnO$_3$ contains Mn$^{3+}$ ions which have one outermost e$_g$ electron. Doping with a divalent alkaline earth ion introduces Mn$^{4+}$ with one  less electron (a hole) and hence manganites with x $<$ 0.5 are called `hole-doped'. Coming from the opposite end of the phase diagram the AMnO$_3$ compound consists of all Mn$^{4+}$ ions and doping it with a trivalent rare earth ion introduces Mn$^{3+}$ ions and hence e$_g$ electrons into the system. Therefore, manganites with x $>$ 0.5 are termed  `electron-doped'. Interestingly, the phase diagram is not symmetric across x = 0.5 concentration though the number of charge carriers on both sides of x = 0.5 vary in a symmetric manner.  Broadly speaking, the x $<$ 0.5 region in the phase diagrams of the manganites is dominated by ferromagnetic interaction whereas x $>$ 0.5 region is characterised by charge ordering.  The phase diagram of Pr$_{1-x}$Ca$_x$MnO$_3$ (PCMO) shows charge ordering in the 0.3 $<$ x $<$ 0.8 composition regime. However, the asymmetry of the phase diagram persists even in the charge ordered (CO) state of x $<$ 0.5 e.g. x=0.36(PCMO-h) and x $>$ 0.5 compound e.g. x=0.64 (PCMO-e). The properties of the CO state are quite different in  the two compounds. The charge ordering transition temperature T$_{co}$ for PCMO-h is 210 K whereas that for PCMO-e is 268 K. Charge ordering in the former can be melted into a ferromagnetic metallic state by doping with ions such as Cr$^{3+}$ and Ru$^{4+}$by the application of magnetic fields whereas no such signs of melting of charge order are observed in PCMO-e. The cause of these differences has been attributed to the intrinsic differences in the electronic structure of the two types of systems\cite{sar}.

Electron Paramagnetic Resonance (EPR) has  proved to be a valuable tool in the study of the manganites.  Through a study of the temperature and composition dependence of various EPR parameters such as the linewidth and the intensity across the various transitions valuable information has been obtained regarding the interplay of different interactions in the systems \cite{ose,riv,cau,tov,she,lof,iva,she1} . However most of  these studies relate to CMR manganites and there are comparatively very few published reports of  EPR studies of CO manganites \cite{raj,jan,dup}.To the best of our knowledge, there has been no report of a comparative study of electron and hole-doped manganites. Since the asymmetry in the phase diagram across x=0.5 is an interesting aspect of the physics of manganites,  we  compare the EPR results on the electron-doped manganite Pr$_{0.36}$Ca$_{0.64}$MnO$_3$ (PCMO-e) with those reported by us  earlier \cite{raj} on the hole-doped Pr$_{0.6}$Ca$_{0.4}$MnO$_3$ (PCMO-h)  with a view to understanding  this asymmetry as reflected in the EPR parameters and their  temperature dependence.

\section{Experimental}

Single crystals of PCMO-e were prepared by the float zone technique.  Resistivity measurements show an increase in the resistivity at T$_{CO}$ = 268 K. Magnetization measurements show a peak in the susceptibility also at 268 K. However no peak in the susceptibility was observed at T$_N$ unlike in PCMO-h.
  
 The EPR experiments were carried out on both single crystal and powder samples of PCMO-e using a Bruker X - band spectrometer (model 200D) equipped with an Oxford Instruments continuous flow cryostat (model ESR 900). The spectrometer was modified by connecting the X and Y inputs of the chart recorder to a 12 bit A/D converter which in turn is connected  to a PC enabling digital data acquisition. With this accessory, for the scanwidth typically used for our experiments i.e 6000 Gauss, one could determine the magnetic field to a precision of  3 Gauss. For single crystal study the static magnetic field was kept parallel to the c-axis of the crystal. Experiments were also performed with another  orientation (H $\parallel $ a) to check for any anisotropy in the ESR response. For measurements on powder, the finely ground sample was dispersed in paraffin wax.The temperature was varied from 4.2 K to room temperature (accuracy: $\pm 1 K$) and the EPR spectra were recorded while warming the sample. Signals could be recorded only for temperatures T $>$ 200 K below which signals had very poor signal to noise ratio.  While doing experiments on both the single crystal and the powder, a speck of DPPH  was used as a g-marker. To carry out the lineshape fitting (to be described below) the signal due to DPPH was subtracted digitally.

\section{Results and Discussion}

Figure 1a shows some representative EPR spectra recorded on the single crystal of PCMO-e. The signals are Dysonian in shape and are fitted to the equation\cite{iva}
$$ {dP\over dH} = {d\over dH}({{\Delta H+\alpha(H-H_0)}\over 4(H-H_0)^2 + \Delta H^2}+ {{\Delta H+\alpha(H+H_0)}\over 4(H+H_0)^2 + \Delta H^2)})\eqno(1)$$ 

where $\Delta$H is the full width at half intensity, H$_0$ is the resonance field and $\alpha$ is the asymmetry parameter which is the fraction of the dispersion component of Lorentzian added to the absorption component resulting in the Dysonian lineshape. In equation 1 the first term represents the signal response due to the  component of microwaves that is polarised clockwise and the second term represents the response  to the component polarised anticlockwise.

Figure 1b shows the EPR spectra from the powder sample of PCMO-e at a few temperatures. These are Lorentzian in shape and are fitted to equation 

$$ {dP\over dH} = {d\over dH}({{\Delta H}\over 4(H-H_0)^2 + \Delta H^2}+ {{\Delta H}\over 4(H+H_0)^2 + \Delta H^2)})\eqno(2)$$ 
where the symbols have the same meanings as in equation 1. The two terms are incorporated for the same reason as mentioned above. The lineshape parameters obtained by fitting are plotted in figures 2, 3 and 4 along with the data of PCMO-h

The  single crystal parameters g, peak to peak linewidth $\Delta$H$_{pp}$(= $\Delta$H / $\sqrt{3}$) and the asymmetry parameter $\alpha$  obtained from the fits to equation 1 are plotted in figure 2 against temperature. The single crystal data of PCMO-h are also plotted in the same figure for the purpose of comparison. 

As can be seen from figure 2a, the  value of `g' (2.002) in  PCMO-h  at room temperature is higher than that in  PCMO-e (1.999) which is less than the free electron  value. The `g' value in  PCMO-e (Fig. 2a, solid squares)  increases only slightly as the temperature is lowered from room temperature to T$_{co}$ below which it decreases sharply. It  reaches a value of 1.94 at 200 K. This temperature dependence is quite different from that observed in the case of PCMO-h (Fig. 2a) where it remains nearly temperature independent in the paramagnetic phase, i.e, from room temperature to T$_{co}$, below which it decreases sharply to a value less than the free electron `g' value, `g$_e$'. With further reduction in temperature, the `g' value increases monotonically.

The linewidth (Fig. 2b) also behaves differently in  the two manganites. In  PCMO-e it shows a slight  monotonic decrease  down to 245 K below which it shows a smooth, sharper increase. In  PCMO-h on the other hand, there is a  sharp discontinuous increase at T$_{co}$ and a monotonic increase at lower temperatures.

The asymmetry parameter $\alpha$ plotted in figure 2c  exhibits  marked differences between the two samples.  While it increases abruptly at T$_{co}$  in  PCMO-h,  in PCMO-e it remains temperature independent till well below T$_{co}$, i.e, down to 245 K and then decreases monotonically below that temperature. In PCMO-h on the other hand throughout the temperature below T$_{co}$, $\alpha$ decreases monotonically. 

In figure 3, we show the results from the powder spectra of PCMO-h and PCMO-e. The `g' values plotted in figure 3a  show a monotonic increase as the temperature is reduced from room temperature in both the compounds. The magnitude of `g'  in  the former is however greater than   `g$_e$'  throughout the temperature range whereas in the latter it shows a crossover from less than `g$_e$' to a  value larger than `g$_e$' near T$_{co}$. Also, the temperature dependence of  `g'  in  the two samples is qualitatively different. The variation of `g'  from room temperature down to T$_N$ is much larger in PCMO-e than in PCMO-h. Another noteworthy point is that the temperature dependences of `g' in the powder and single crystal of PCMO-e are  opposite to each other. Moreno et al., \cite{mor} observed an anisotropy in the temperature dependence of `g' in single crystals of layered manganite La$_{2(1-x)}$Sr$_{1+2x}$Mn$_2$O$_7$. They observed that on decreasing the temperature  `g' increases with T for H$\parallel$a but decreases with T for H$\parallel$c. In order to examine the possibility of such an anisotropy in the single crystal of PCMO-e, we repeated the experiment for  H$\parallel$a and found that there was no anisotropy in the variation of `g' with temperature. So  the reasons for the difference between behaviour of `g'  in the single crystal and the powder of PCMO-e are not clear at present. 

The  dependence of the peak to peak linewidth $\Delta$H$_{pp}$ on temperature shown in figure 3b  for PCMO-h and PCMO-e also shows a difference  below T$_{co}$ in that the curvatures of the two are opposite to each other. The variation in linewidth is again larger in PCMO-e than in PCMO-h.

The EPR intensities for the two samples are plotted in figures 3c  against temperature. In  both the samples, the intensity goes through a peak at T$_{co}$, decreasing monotonically below this temperature. However, the sharp drop in intensity found at $\sim$ T$_{co}$ (240 K) in the case of PCMO-h is not seen  in PCMO-e. 

Now we attempt to understand these results in the light of the microscopic phenomena occurring in these systems. We focus on the temperature variation of `g'  in the powder sample and that of  $\Delta$H and $\alpha$ in single crystals. 
It was earlier observed  by us that the hole-doped charge ordering systems exhibit certain tell-tale EPR signatures: 
\begin {enumerate}
\item It was noted that  `g' ($g = g_e(1 \pm k{\lambda \over \Delta})$ where $\lambda$ is the spin-orbit coupling constant, $k$ is a positive numerical factor and $\Delta$ is the crystal field splitting), even in the charge disordered paramagnetic state, is larger than `g$_e$' and on cooling stays independent of temperature till T$_{co}$. Below T$_{co}$ it becomes very sensitive to temperature and goes on increasing with decreasing temperature. The anomalous  `g' shift (since for both Mn$^{3+}$  and Mn$^{4+}$ `g' is expected to be lower than `g$_e$')  can be  attributed  to the `hole' nature  of the charge carriers since for holes in a less than half filled shell the spin-orbit coupling constant $\lambda $ is  negative \cite{wel}. This conclusion is consistent with the present observation that in the system with electrons as charge carriers, `g' is less than `g$_e$' in the charge disordered state. The increase in `g' below T$_{co}$ found in PCMO-h and NdCaMnO$_3$ (NCMO)\cite{jan} was attributed to the strengthening of the spin-orbit interaction due to orbital ordering developing  between T$_{co}$ and T$_N$. However, the fact that `g' crosses over to a value greater than `g$_e$' in the electron-doped system and continues to increase as the temperature is lowered below T$_{co}$ indicates that an essentially different mechanism is operative in the electron-doped system. This is consistent with the suggestion by Khomskii\cite{kho}that in these systems the concept of orbital ordering cannot be strictly applied. 
\item In the hole-doped PCMO and NCMO  the  dependence of $\Delta$H on T in the single crystals in the region T$_N$ $\le$ T$_{co}$ could be explained in terms  of a model involving motional narrowing\cite{jan}. Hopping of the Jan-Teller polarons was understood to be  the underlying mechanism of charge transport. However, we notice in Figs. 2b,  3b and 4 that $\Delta$H {\it vs} T of PCMO-e and PCMO-h have different functional dependences  on temperature. 
Interestingly we observe that the temperature dependence of $\rho$  is the same as the temperature dependence of  $\Delta$H$_{pp}$ of single crystal  as shown in the inset of Fig. 4. We find that  both the resistivity and the single crystal EPR linewidth fit the model of variable range hopping (VRH)(solid line in Fig 4) given by equation $$\Delta H = K. exp{({T_0\over T})^{1\over4}}\eqno(3)$$ VRH has been used to explain the resistivity of manganites in the paramagnetic state by Viret et al., \cite{vir} using the concept of magnetic localisation. They found that the density of states gets modified consequent to the localisation of charge carriers due to Hund's  coupling between the itinerant e$_g$ electrons and the stationary, core t$_{2g}$ spins. This lead to the realisation of a random potential in the paramagnetic state. In case of CO manganites, in the charge ordered state a long ranged AFM order is not established. So the concept of random magnetic potential can be applied. From the VRH fit we obtained the parameter T$_0$. The localisation length is given by ${1\over{\alpha}} = \sqrt[3] {k.T_0\over{171.U_m.v}} $ where k is the Boltzmann factor, $U_m$ is the random magnetic potential which is of the order of 2 eV (Hund's coupling) and $v$ is the volume of the unit cell per manganese ion which is $5.7 X 10^{-29} m^3$. The numerical factor 171 comes from the geometrical corrections due to the symmetry of $d_{z^2}$ orbital of e$_g$ electron and the probability that an unoccupied manganese orbital can accept an electron.  The localisation length for PCMO-e thus obtained is  0.3 nm which is larger than the Mn ionic radius ($\sim$ 0.8 A$^o$) as expected in VRH model.  While the VRH fit to $\Delta$H$_{pp}$ {\it vs} T in PCMO-e is seen to be excellent, the model did not fit the linewidth data of PCMO-h well (dotted line in figure 4), the regression coefficient indicative of the goodness of fit being 0.91 in the latter compared to 0.99 for the fit to the data of PCMO-e. 
\item The same conclusion i.e. the applicability of the model of motional narrowing in the case of PCMO-e is borne out by the behaviour of $\alpha$ as well. In case of PCMO-h an anomalous peak is observed in $\alpha$ {\it vs} T at T$_{co}$. No such peak is observed in PCMO-e where $\alpha$ decreases smoothly as $\rho$ increases below T$_{co}$ as expected. In PCMO-h, the peak in $\alpha$ was found to be correlated with the motion of the Jahn-Teller polarons. The absence of such a peak in PCMO-e again points towards the negligible contribution of Jahn-Teller polaron-mediated  charge transport  in this system. 
\end {enumerate}

In summary, the comparative EPR study of hole-doped and electron-doped PCMO brings out certain essential differences in the nature of the order and the transport mechanisms in the two cases.

{ \bf Acknowledgments}

AKS and  SVB thank Department of Science and Technology for financial support. JJ thanks CSIR, India for a fellowship. CNRR thanks the BRNS (DAE) for support.

\newpage
\begin{large}
\noindent {Figure Captions}\\
\end{large}
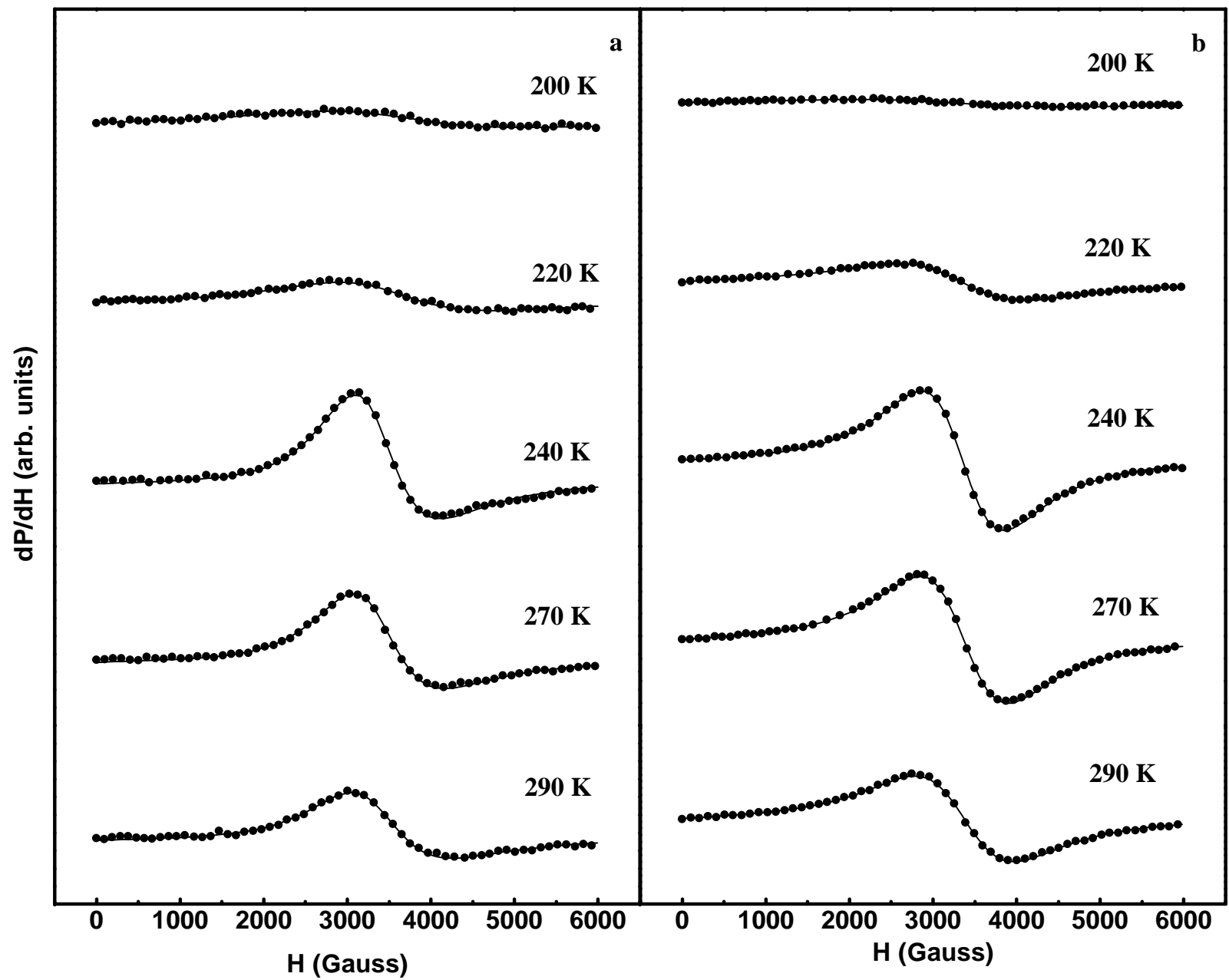
\noindent { FIGURE 1}

 (a) EPR spectra of a single crystal sample of Pr$_{0.36}$Ca$_{0.64}$MnO$_3$ for a few representative temperatures. The solid circles are experimental data and the solid line shows the fit to  equation 1 described in the text. (b) EPR spectra of a powder sample of Pr$_{0.36}$Ca$_{0.64}$MnO$_3$ for a few representative temperatures. The solid circles are experimental data and the solid line shows the fit to  equation 2 described in the text. \\ 
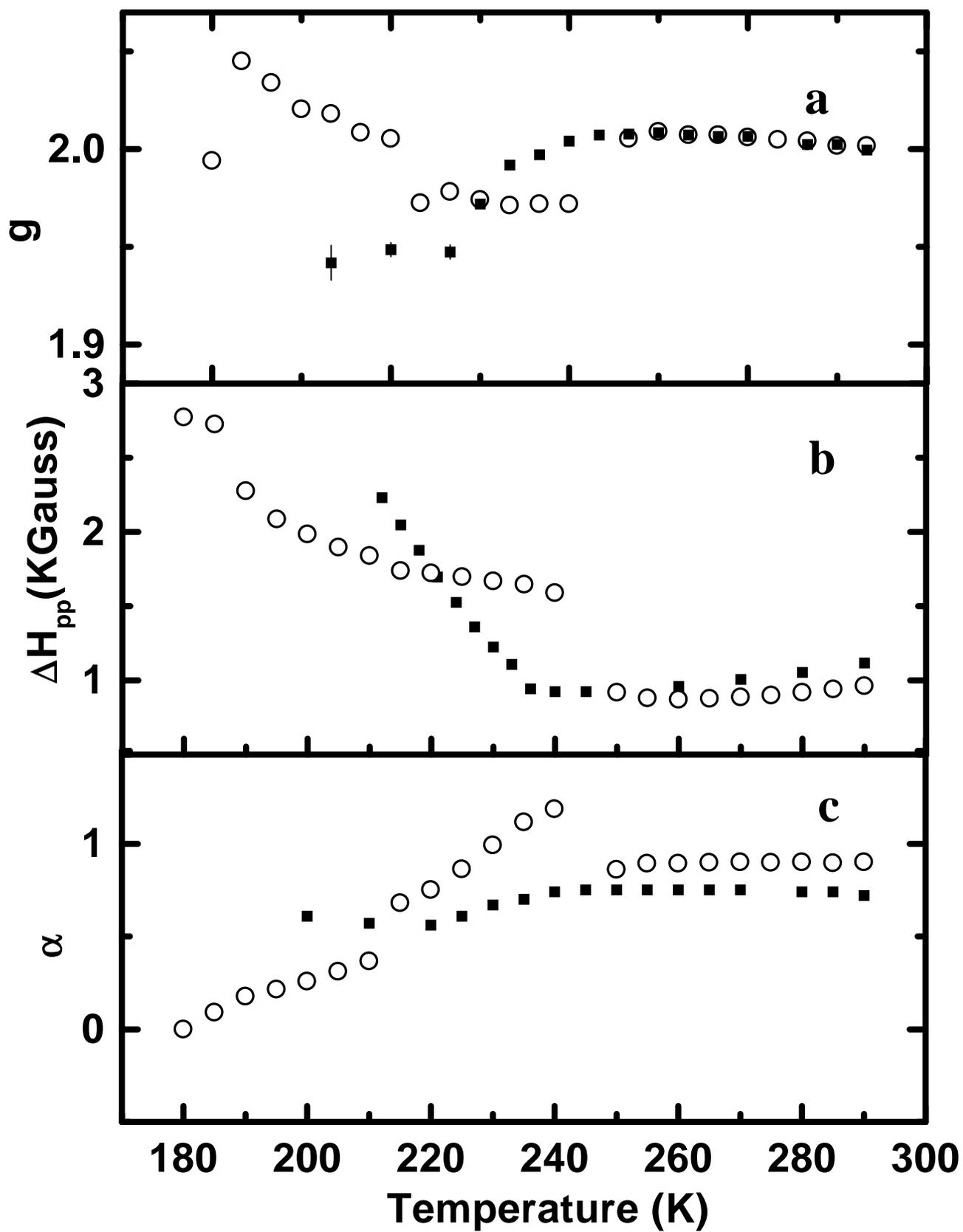
\noindent {FIGURE 2}

Temperature dependence of single crystal lineshape parameters obtained from the fits to equation 1.(a) g; the vertical lines are error bars (b) peak to peak linewidths $\Delta$H$_{pp}$ and (c) $\alpha$; the error bars for $\Delta$H$_{pp}$ and $\alpha$ are very small and hence are not shown . The open circles represent the data for PCMO-h and the solid squares for PCMO-e.\\
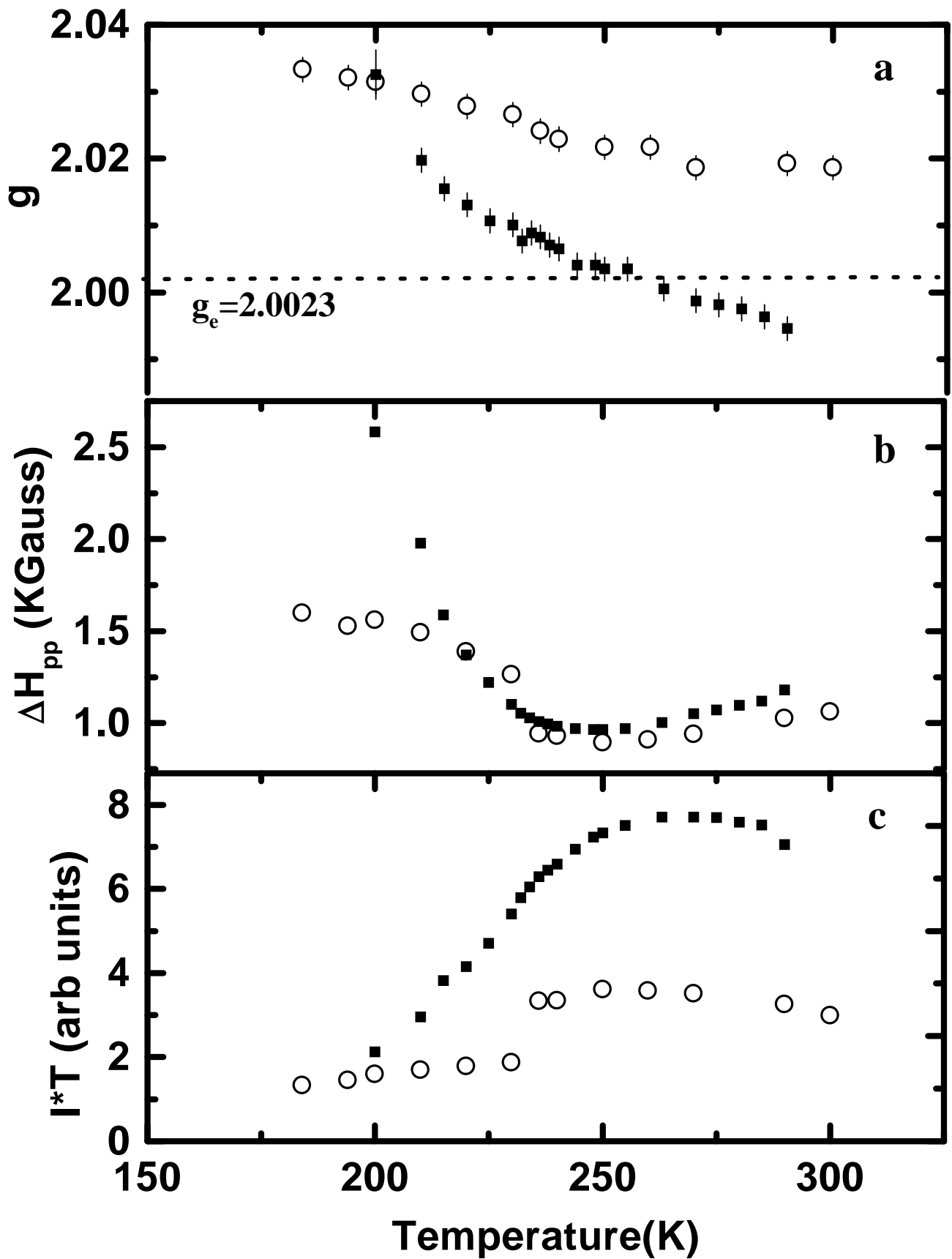
\noindent {FIGURE 3}

Temperature dependence of powder lineshape parameters obtained from the fits to equation 2. (a) g; The vertical lines show error bars on the `g' value (b) the peak to peak linewidths $\Delta$H$_{pp}$   and  (c) Intensity times T. The error bars for the $\Delta$H$_{pp}$ and intensity  are very small and hence are not shown . The open circles represent the data for PCMO-h and the solid squares for PCMO-e.\\
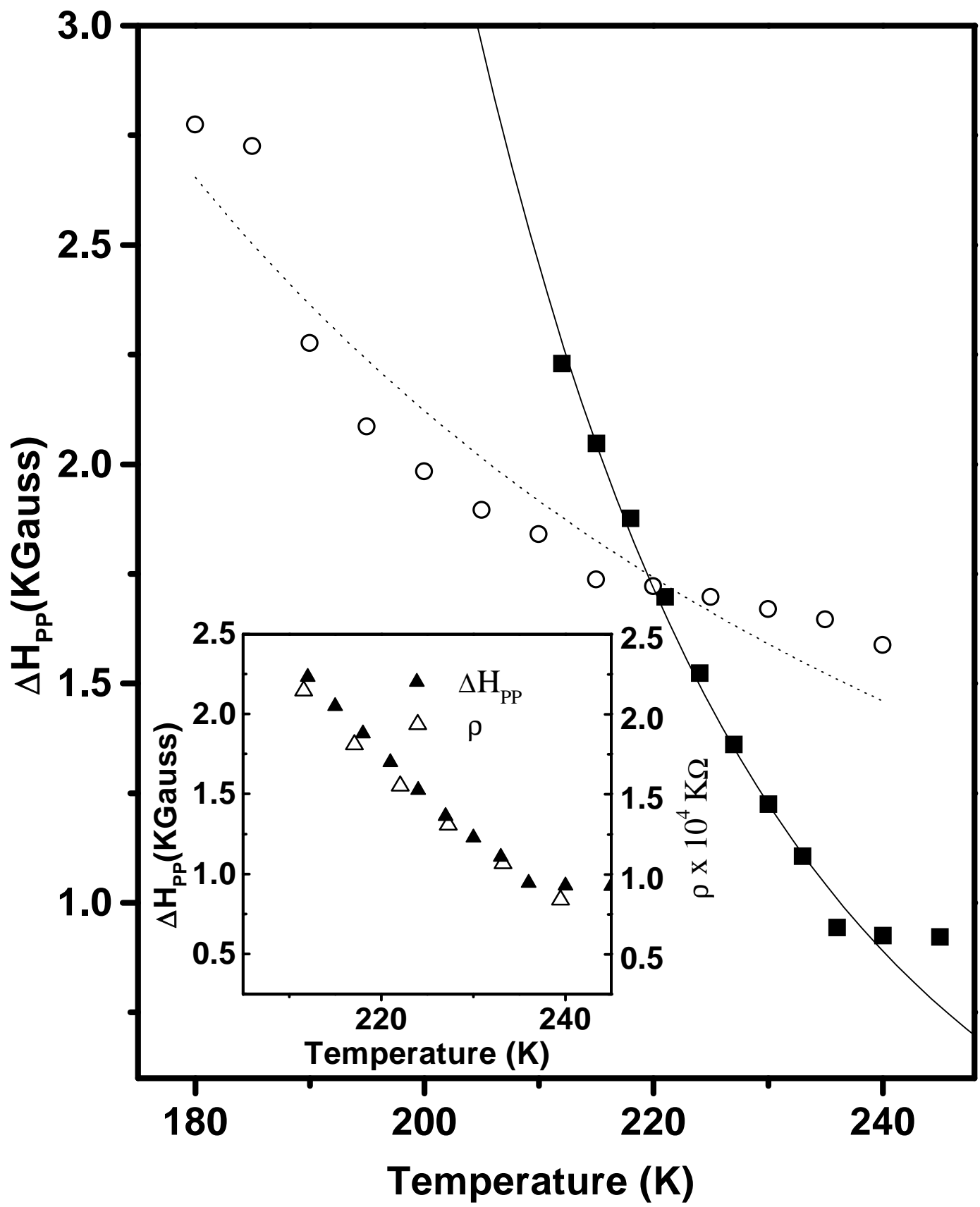
\noindent {FIGURE 4}

Temperature dependence of single crystal linewidths $\Delta$H$_{pp}$ obtained from the fits to equation 1. The solid squares are the data for PCMO-e and the open circles for PCMO-h. The solid and the dotted lines are the least square fits to the variable range hopping model (Eqn. 3) to the respective data sets in the temperature range of T$_N$ to  T$_{co}$.  The inset  shows the essentially similar temperature dependence of the resistivity (obtained from reference 3, scaled by an appropriate  constant,   open triangles), and  $\Delta$H$_{pp}$ (solid triangles) data for PCMO-e.

\end{document}